\documentclass[12pt]{article}

\usepackage[dvipdfmx]{graphicx}
\usepackage{amsfonts}
\usepackage[mathscr]{eucal}
\usepackage{ascmac}
\usepackage{dcolumn}
\usepackage{bm}
\usepackage[colorlinks=true,linkcolor=blue,citecolor=blue]{hyperref}
\usepackage{hyperref}
\usepackage{color}

\begin{document}


\newcommand{\gtrsim}{ \mathop{}_{\textstyle \sim}^{\textstyle >} }
\newcommand{\lesssim}{ \mathop{}_{\textstyle \sim}^{\textstyle <} }
\newcommand{\vev}[1]{ \left\langle {#1} \right\rangle }
\newcommand{\bra}[1]{ \langle {#1} | }
\newcommand{\ket}[1]{ | {#1} \rangle }
\newcommand{\EV}{ \ {\rm eV} }
\newcommand{\KEV}{ \ {\rm keV} }
\newcommand{\MEV}{\  {\rm MeV} }
\newcommand{\GEV}{\  {\rm GeV} }
\newcommand{\TEV}{\  {\rm TeV} }
\newcommand{\1}{\mbox{1}\hspace{-0.25em}\mbox{l}}
\newcommand{\Red}[1]{{\color{red} {#1}}}

\newcommand{\lmk}{\left(}  
\newcommand{\rmk}{\right)}
\newcommand{\lkk}{\left[}  
\newcommand{\rkk}{\right]}
\newcommand{\lhk}{\left \{ }  
\newcommand{\rhk}{\right \} }
\newcommand{\del}{\partial}  
\newcommand{\la}{\left\langle} 
\newcommand{\ra}{\right\rangle}
\newcommand{\half}{\frac{1}{2}}

\newcommand{\bea}{\begin{array}}
\newcommand{\eea}{\end{array}}
\newcommand{\beq}{\begin{eqnarray}}
\newcommand{\eeq}{\end{eqnarray}}

\newcommand{\dd}{\mathrm{d}}
\newcommand{\Mpl}{M_{\rm Pl}}
\newcommand{\mg}{m_{3/2}}
\newcommand{\abs}[1]{\left\vert {#1} \right\vert}
\newcommand{\mphi}{m_{\phi}}
\newcommand{\Hz}{\ {\rm Hz}}
\newcommand{\for}{\quad \text{for }}
\newcommand{\Min}{\text{Min}}
\newcommand{\Max}{\text{Max}}
\newcommand{\Kahler}{K\"{a}hler }
\newcommand{\cphi}{\varphi}
\newcommand{\Tr}{\text{Tr}}
\newcommand{\diag}{\text{diag}}

\newcommand{\SUf}{SU(3)_{\rm f}}
\newcommand{\Upq}{U(1)_{\rm PQ}}
\newcommand{\Zpq}{Z^{\rm PQ}_3}
\newcommand{\Cpq}{C_{\rm PQ}}
\newcommand{\ubar}{u^c}
\newcommand{\dbar}{d^c}
\newcommand{\ebar}{e^c}
\newcommand{\nubar}{\nu^c}
\newcommand{\Ndw}{N_{\rm DW}}
\newcommand{\Fpq}{F_{\rm PQ}}
\newcommand{\fpq}{v_{\rm PQ}}
\newcommand{\Br}{{\rm Br}}
\newcommand{\Lag}{\mathcal{L}}
\newcommand{\Lqcd}{\Lambda_{\rm QCD}}
\newcommand{\eq}[1]{Eq.~(\ref{#1})}


\begin{titlepage}

\baselineskip 8mm

\begin{flushright}
IPMU 15-0210
\end{flushright}

\begin{center}

\vskip 1.2cm

{\Large\bf
High-scale SUSY 
from~an~R-invariant~New~Inflation in~the~Landscape
}

\vskip 1.8cm

{\large 
Masahiro Kawasaki$^{a,b}$, 
Masaki Yamada$^{a,b}$, 
Tsutomu T. Yanagida$^{b,c}$, 
\\
and 
Norimi Yokozaki$^{d}$,
}

\vskip 0.4cm

{\it$^a$
ICRR, 
The University of Tokyo,
Kashiwa, Chiba 277-8582, Japan}\\
{\it$^b$Kavli IPMU (WPI), UTIAS, The University of Tokyo, 
Kashiwa, Chiba 277-8583, Japan}\\
{\it$^c$
ARC Centre of Excellence for Particle Physics at the Terascale,
School of Physics, The University of Melbourne,
Victoria 3010, Australia}\\
{\it$^d$
INFN, 
Sezione di Roma, 
Piazzale Adlo Moro 2, I-00185 Rome, Italy}

\date{\today}
\vspace{2cm}

\begin{abstract}  
We provide an anthropic reason that 
the supersymmetry breaking scale is much higher than the electroweak scale 
as indicated by the null result of collider experiments and observed $125 \GEV$ Higgs boson. 
We focus on a new inflation model 
as a typical low-scale inflation model that may be expected in the string landscape. 
In this model, the R-symmetry is broken at the minimum of the inflaton potential 
and its breaking scale is related to the reheating temperature. 
Once we admit that 
the anthropic principle requires 
thermal leptogenesis, 
we obtain a lower bound on 
gravitino mass, which is related to R-symmetry breaking scale. 
This scenario and resulting gravitino mass 
predict the consistent amplitude of density perturbations. 
We also find that string axions and saxions are consistently implemented in this scenario.

\end{abstract}


\end{center}
\end{titlepage}

\baselineskip 6mm


\section{Introduction
\label{sec:introduction}}

The observed $125 \GEV$ Higgs boson~\cite{Aad:2012tfa, Chatrchyan:2012ufa}
and the null result of supersymmetric (SUSY) particles at the LHC 
may imply that the SUSY breaking scale is much higher than the electroweak scale~\cite{
Okada:1990vk, Ellis:1990nz, Haber:1990aw}. 
This seems to be unnatural in light of the (little) hierarchy problem, 
which would compel one to search for the reason 
that the nature ``selects'' such a large SUSY breaking scale. 
In this paper 
we provide its cosmological reason in the anthropic landscape.

The string landscape 
indicates that there are a lot of local vacua with different values of potential energy, 
which implies that 
universes can have any values of cosmological constant~\cite{Bousso:2000xa, Kachru:2003aw, Susskind:2003kw, Douglas:2003um}. 
In the anthropic landscape, 
it is argued that 
we cannot live in 
a universe with a larger cosmological constant than the observed value or with a negative one, 
so that our universe is the one with a marginal value of cosmological constant, 
which is consistent with the observation~\cite{Weinberg:1987dv} 
(see also Ref.~\cite{Hogan:1999wh, Hall:2007ja}). 
This may be a unique solution to the cosmological constant problem 
in our present understanding, 
so that in this paper we stand on this scenario.

Low-scale inflation may be a natural consequence of the anthropic landscape, 
where inflations occur at infinitely many vacua. 
Old inflation occurs at a local vacuum 
and it ends via the tunnelling to a vacuum with a smaller vacuum energy. 
Then inflation may continue at that vacuum. 
Eventually, 
the inflaton tunnels into a vacuum 
around which 
observable slow-roll inflation occurs. 
This scenario implies that 
the observable inflation is that with a relatively small energy scale. 
Among simple inflation models in supersymmetry (SUSY), 
including chaotic inflation~\cite{Kawasaki:2000yn, Kallosh:2010ug} 
and hybrid inflation models~\cite{Copeland:1994vg, Dvali:1994ms}, 
new inflation models are one class of the simplest and smallest-energy scale inflation models~\cite{Izawa:1996dv, 
Nakayama:2012dw, Takahashi:2013cxa, Harigaya:2013pla}. 
Therefore, in this paper, 
we focus on 
the new inflation model considered in Refs.~\cite{Izawa:1996dv, Harigaya:2013pla}. 
In this model, 
R-symmetry is broken at the global minimum 
and gravitino mass depends on parameters in the inflaton sector. 
As a result, the upper-bound on reheating temperature is related to the gravitino mass. 
We find that 
the gravitino mass needs to be larger than of order $100 \TEV$ 
in order to realize the thermal leptogenesis~\cite{Fukugita:1986hr}, 
which may be required by the anthropic principle. 
This is the reason that 
the SUSY breaking scale is much higher than the electroweak scale 
as indicated by the collider experiments.

The gravitino with mass of $100 \TEV$ 
is consistent with the other observations and constraints in the anthropic landscape. 
First, 
the observed amplitude of density perturbations is a natural consequence 
of our model with $100 \TEV$ gravitino mass. 
Secondly, 
string axions~\cite{Peccei:1977hh, Weinberg:1977ma, Wilczek:1977pj, 
Conlon:2006tq, Svrcek:2006yi, Choi:2006za}, 
whose decay constants are of order the grand unified theory (GUT) scale, 
can be consistently introduced in our scenario. 
Their initial amplitudes 
are fine-tuned not to overclose the Universe in the anthropic landscape, 
while 
the constraint on axion isocurvature perturbation is avoided due to the low-energy nature of new inflation~\cite{Linde:1984ti,Seckel:1985tj,Lyth:1989pb,Turner:1990uz,Linde:1991km, Ade:2015lrj}. 
In the anthropic landscape, 
the lightest SUSY particle (LSP) overproduction problem by saxion decay is also avoided by fine-tunings of saxion initial amplitudes, 
where we assume R-parity conservation. 
This implies that 
the saxion decay does not produce much entropy, 
so that the thermal leptogenesis can be realized consistently. 
The saxion, whose mass is about gravitino mass of order $100 \TEV$, 
safely decays into radiation before the Big Bang nucleosynthesis (BBN) epoch, which is required 
by the consistency of our scenario with the BBN theory~\cite{Kawasaki:1999na}. 
Finally, the gravitino problem is also alleviated by $100 \TEV$ gravitino mass 
because 
such a heavy gravitino safely decays before the BBN epoch. 
The thermal relic of the LSP is not overabundant 
when the LSP is wino or higgsino with a mass less than $\mathcal{O}(1) \TEV$~\cite{Ibe:2012hu} 
(see also Refs.~\cite{Hall:2012zp}),%
\footnote{
See an early work~\cite{Wells:2004di}. 
}
which is naturally realized in pure gravity mediation~\cite{puregm, Ibe:2011aa}.%
\footnote{
For a similar model, see Ref.~\cite{Hall:2011jd}. 
}
In this case, 
the produced LSP from gravitino decay 
does not overclose the Universe 
for reheating temperature of order $10^9 \GEV$.

\paragraph{}

Here we summarize our standing point in this paper. 
\begin{enumerate}
\item Anthropic landscape: 

The cosmological constant problem is solved by the anthropic principle. 
We consider a new inflation model that may be favoured in the string landscape.

\item Thermal leptogenesis: 

The observed baryon asymmetry is generated by the thermal leptogenesis. 
As a result, $100 \TEV$ gravitino mass is predicted in the anthropic landscape. 

\item String axions: 

The strong CP problem is solved by the PQ mechanism. 
The dark matter (DM) overproduction problems from axion oscillations and saxion decays 
are avoided by fine-tunings of their initial amplitude in the anthropic landscape.

\end{enumerate}

This paper is organized as follows. 
In the next section, 
we consider the new inflation model 
with R-symmetry breaking at the potential minimum. 
We discuss the reheating process 
and explain that 
the thermal leptogenesis 
can be realized only for gravitino mass of order $100 \TEV$. 
Then we derive a condition to the realization of eternal inflation 
in new inflation, 
which may provide a reason of smallness of a parameter in \Kahler potential. 
In Sec.~\ref{sec3}, 
we explain axion and saxion cosmology 
and discuss consistency of our scenario 
with string axions. 
Section~\ref{sec:conclusion} is devoted to conclusion.

\section{New inflation 
\label{sec:sec2}}

If many inflations occur in the string landscape, 
it is natural to consider that the observable inflation is 
the one with the smallest energy scale. 
Thus we consider a new inflation model, 
which is 
one of the simplest and smallest-energy scale inflation models 
among inflation models in SUSY.

\subsection{Model
\label{model}}

We consider the new inflation model in supergravity 
that is realized by the following superpotential~\cite{Izawa:1996dv}: 
\beq 
 W = v^2 \phi - \frac{g}{6} \phi^{6}, 
\eeq
where $v$ and $g$ are parameters 
and can be taken to be real positive by $U(1)_R$ rotation and field redefinition 
without loss of generality. 
Here we implicitly assume $Z_{10 R}$ symmetry.%
\footnote{
A model with $Z_{6R}$ symmetry predicts too red-tilted spectral index compared with the observed value~\cite{
Harigaya:2013pla}. 
It is an open question why the nature respects $Z_{10 R}$ symmetry. 
}
We use the Planck unit where the reduced Planck scale ($\simeq 2.4 \times 10^{18} \GEV$) 
is identified with unity. 
We consider a \Kahler potential of 
\beq
 K = \abs{\phi}^2 + \frac{k}{4} \abs{\phi}^4, 
 \label{Kahler}
\eeq
where $k$ is a real parameter. 
One might naively expect that $k$ is of order unity, 
but for a moment we take it as a free parameter. 
In supergravity, 
the potential of $\phi$ is written as 
\beq
 V_{\rm new} = v^4 - k v^4 \abs{\phi}^2 - g v^2 \lmk \phi^5 + \phi^{*5} \rmk 
 + g^2 \abs{\phi}^{10} 
 + \dots, 
\eeq
where we assume $v^2 \ll 1$ and neglect higher-dimensional terms. 
At the minimum of the potential, 
the VEV of the inflaton $\phi$ is given by 
\beq
 \la \phi \ra_{\rm min} 
 \simeq \lmk \frac{v^2}{g} \rmk^{1/5}, 
 \label{VEV}
\eeq
and 
the $Z_{10R}$ symmetry is broken down to the R-parity $Z_{2R}$. 
Note that SUSY is not broken at the minimum of the potential 
while 
the superpotential has a nonzero value of 
\beq
 \la W \ra_{\rm min} 
 \simeq 
 \frac{5}{6} v^2 \lmk \frac{v^2}{g} \rmk^{1/5}, 
\eeq
so that gravitino obtains a mass of $m_{3/2} = e^{\la K \ra / 2} \la W \ra_{\rm min} \simeq \la W \ra_{\rm min} $.

The properties and predictions of this new inflation model has been 
investigated in detail in Ref.~\cite{Harigaya:2013pla}. 
Here we quote their results. 
The spectral index can be within the observed value 
(i.e., $n_s = [0.952, 0.976]$) 
for 
\beq
 k = [10^{-2.6}, 10^{-1.7}]. 
 \label{n_s}
\eeq
We obtain 
the observed amplitude of the curvature perturbations $\mathcal{P}_\zeta \simeq 2.2 \times 10^{-9}$ 
when $v$ is given by 
\beq
 v 
 \simeq 1.1 \times 10^{13} \GEV \lmk \frac{g}{1000} \rmk^{-1/4}, 
 \label{v}
\eeq
where we assume $k=0.01$ and the $e$-folding number to be $50$. 
In the next section, we predict $v$ of order this value 
in the anthropic landscape. 
Equation~(\ref{v}) implies the energy scale of inflation as 
\beq 
 H_{\rm new} \approx 10^7 \GEV, 
\eeq
and the gravitino mass of 
\beq
 m_{3/2} \simeq 
 \frac{5 g}{6} \lmk \frac{v^2}{g} \rmk^{6/5} 
 \simeq 
 8.0 \times 10^4 \GEV \lmk \frac{g}{1000} \rmk^{-4/5}. 
 \label{m_3/2}
\eeq
We expect that 
the gravitino mass should be as small as possible 
so that fine-tunings of electroweak scale and cosmological constant 
are as small as possible. 
Thus we take the largest possible $g$. 
A large coupling constant $g$ leads to a large radiative correction to the \Kahler potential. 
Counting the number of loops and symmetry factors, 
we estimate the radiative correction such as 
\beq
 \delta K \sim 
 \frac{5!}{(16 \pi^2)^4} g^2 \Mpl^6 \abs{\phi}^2 
 + 
 \frac{4!}{(16 \pi^2)^3} g^2 \Mpl^4 \abs{\phi}^4  
 + \dots, 
\eeq
where we assume the cutoff scale to be the Planck scale 
and 
explicitly write it as $\Mpl$. 
The unitarity limit $\abs{\delta K} \lesssim \abs{\phi}^2$ 
leads to an upper bound on $g$ as $g \lesssim 2000$. 
Note that this is consistent with the naive dimensional analysis. 
Hereafter we take $g \sim 1000$ 
and 
then the gravitino mass is of order $100 \TEV$. 
Note that the parameter $k$ in the \Kahler potential of Eq.~(\ref{Kahler}) 
is as large as $100$ due to the radiative correction. 
This implies that 
there is a fine-tuning of order $0.01\%$ in the \Kahler potential 
for the spectral index to be consistent with the observed value. 
In Sec.~\ref{eternal inf}, we provide a scenario to explain this fine-tuning.

\subsection{Thermal leptogenesis and high-scale SUSY}

Here we discuss the reheating process after the new inflation. 
We introduce additional GUT multiplets $Q$ and $\bar{Q}$ 
and 
assign R-charges and $U(1)_{B-L}$ charges as shown in Table~\ref{table1}, 
where $\bm{5}^*$ and $\bm{10}$ are Minimal SUSY Standard Model (MSSM) GUT multiplets 
and $N$ is the right-handed neutrino. 
Then 
we can write the superpotential of 
\beq
 W_{\rm decay} = \frac{y}{3} \phi^3 Q \bar{Q} + y' {\bar Q} {\rm 10} H_d. 
\eeq
The first term leads to the decay of the inflaton into $Q$ and $\bar{Q}$ 
while the second one allows $Q$ and $\bar{Q}$ to decay into MSSM particles subsequently. 
Note that since the potential minimum is found to be Eq.~(\ref{VEV}) and $\la Q \ra = \la \bar{Q} \ra = 0$ 
in this model, 
the above superpotential does not affect the scenario of the new inflation.

\begin{table}\begin{center}
\begin{tabular}{|c|p{1.0cm}|p{1.0cm}|p{1.0cm}|p{1.0cm}|p{1.0cm}|p{1.0cm}|p{1.0cm}|p{1.2cm}|p{1.2cm}|}
  \hline
    & \hfil $\bm{5}^*$ \hfil & \hfil $\bm{10}$ \hfil & \hfil $H_u$ \hfil & \hfil $H_d$ \hfil & \hfil $N$ \hfil & \hfil $\phi$  \hfil & \hfil $Q (\bm{5})$ \hfil & \hfil $\bar{Q} (\bm{5}^*)$\hfil  \\
  \hline
  \hfil $Z_{10R}$ \hfil & \hfil 1 \hfil & \hfil 1 \hfil & \hfil 0\hfil & \hfil 0 \hfil & \hfil 1 \hfil & \hfil 2 \hfil & \hfil 5 \hfil & \hfil 1 \hfil  \\
  \hline
  \hfil $U(1)_{B-L}$ \hfil & \hfil $-3$ \hfil & \hfil $1$ \hfil & \hfil $-2$ \hfil & \hfil $2$ \hfil & \hfil $5$ \hfil & \hfil $0$ \hfil & \hfil $3$ \hfil & \hfil $-3$ \hfil  \\
\hline
\end{tabular}\end{center}
\caption{Charge assignment for matter fields.
\label{table1}}
\end{table}

The effective mass of $Q$ and $\bar{Q}$ at the global minimum is given by 
\beq
 m_Q 
 &=& \frac{y}{3} \la \phi \ra^3 
 \\
 &\simeq& \frac{y}{3} \lmk \frac{v^2}{g} \rmk^{3/5}, 
\eeq
while 
the mass of the inflaton is given by 
\beq
 m_\phi \simeq 5 g \lmk \frac{v^2}{g} \rmk^{4/5}. 
\eeq
We require $m_\phi \ge 2 m_Q$ so that the inflaton can decay into $Q$ and $\bar{Q}$. 
The decay rate is then estimated as 
\beq
 \Gamma_\phi 
 &\simeq& 
 \frac{5}{8 \pi} y^2 \la \phi \ra^4 m_\phi 
 \\
 &\simeq& 
 \frac{25 g}{8 \pi} y^2  
 \lmk \frac{v^2}{g} \rmk^{8/5}, 
\eeq
which leads to the reheating temperature given by 
\beq
 T_{\rm RH} 
 &\simeq& 
 \lmk  \sqrt{\frac{90}{g_* \pi^2}} \Gamma_\phi \Mpl \rmk^{1/2} 
 \\
 &\simeq& 10^9 \GEV 
 \lmk \frac{y}{10} \rmk
 \lmk \frac{v}{5 \times 10^{12} \GEV} \rmk^{8/5} 
 \lmk \frac{g}{1000} \rmk^{-3/10}. 
\eeq

\begin{figure}[t]
\centering 
\includegraphics[width=.40\textwidth, bb=0 0 360 339]{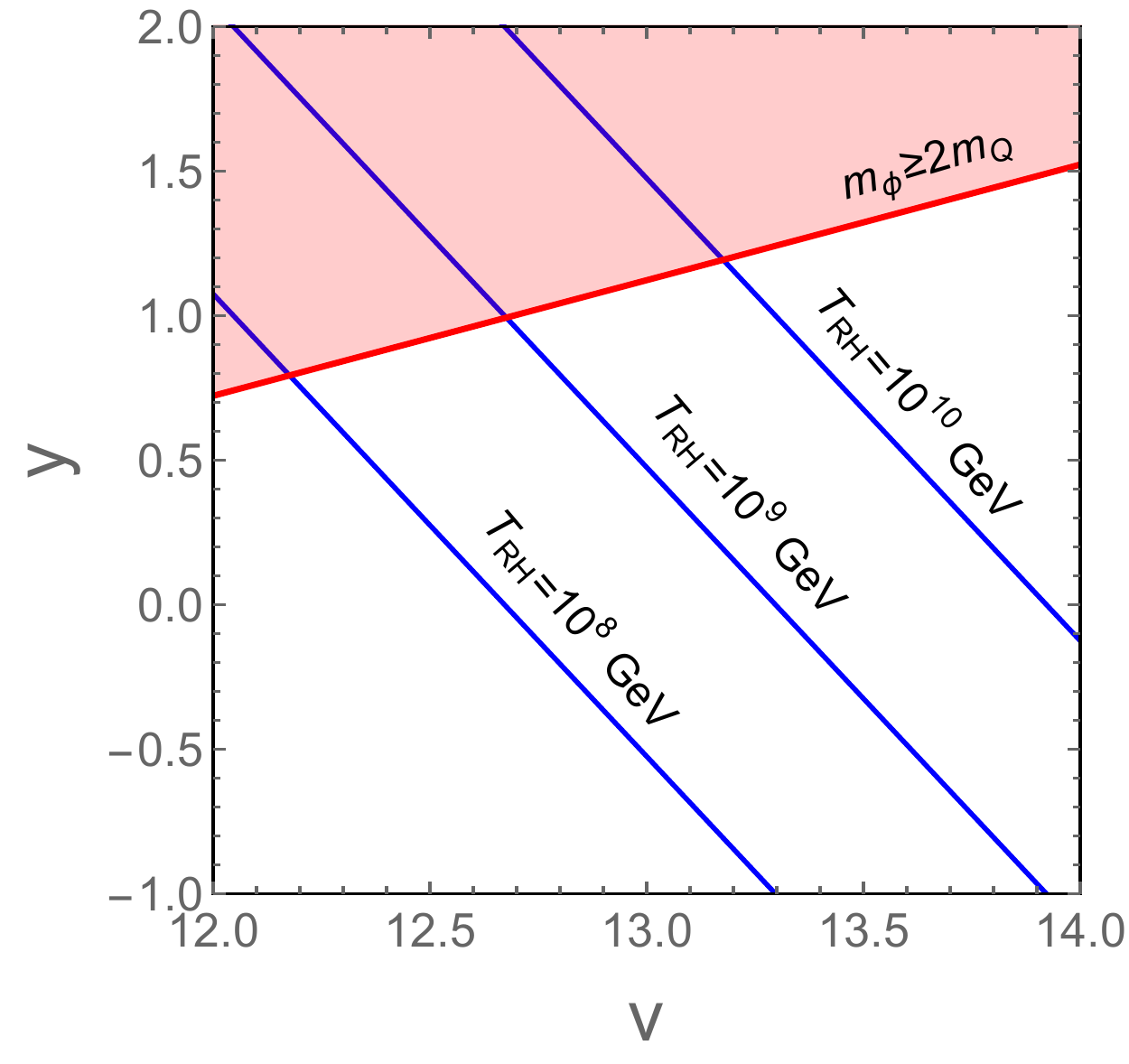} 
\caption{
Contours of reheating temperature in the $v$-$y$ plane. 
The inflaton cannot decay into $Q$ and $\bar{Q}$ kinematically 
in the red-shaded region. 
We set $g=1000$. 
}
  \label{fig1}
\end{figure}

We use the anthropic principle 
to explain the observed amount of baryon asymmetry. 
We assume that 
the baryon asymmetry is generated by the thermal leptogenesis, 
which requires a reheating temperature larger than of order $10^{9} \GEV$~\cite{Fukugita:1986hr, Buchmuller:2005eh}. 
We plot the contours of reheating temperature in $v$-$y$ plane with $g = 1000$ in Fig.~\ref{fig1}, 
where the inflaton cannot decay into $Q$ and $\bar{Q}$ kinematically 
in the red-shaded region. 
Since lighter gravitino mass is favoured from viewpoint of fine-tunings of electroweak scale 
and cosmological constant, 
smaller $v$ should be selected. 
Thus 
we find that $T_{\rm RH} \gtrsim 10^9 \GEV$ and $m_\phi \ge 2 m_Q$ imply 
\beq 
 v &\simeq& 5 \times 10^{12} \GEV 
 \\
 y &\simeq& 10. 
\eeq
It is also consistent with the large value of $g$. 
From Eqs.~(\ref{v}) and (\ref{m_3/2}), 
the amplitude of curvature perturbations is predicted to be of order the observed value 
and the gravitino mass of order $100 \TEV$.

Here we comment on resonant leptogenesis~\cite{Covi:1996wh, Pilaftsis:1997dr}. 
If the Yukawa coupling constants for interactions between the right-handed neutrinos 
and a $B-L$ breaking field can be fine-tuned, 
the right-handed neutrino mass can be degenerated. 
In this case, 
the CP violating effect on their decay 
can be enhanced by a resonance effect 
and 
the lower bound on reheating temperature 
can be smaller than $10^{9} \GEV$ to explain the observed amount of baryon asymmetry. 
Here, 
we provide a conjecture that 
the Yukawa coupling constants are fixed by a mechanism beyond 
the present scope of the string landscape. 
This is motivated by the following discussion. 
In the SM sector, 
if the Yukawa coupling constants are not fixed, 
the Higgs VEV may not be determined by the anthropic principle. 
In fact, 
the Higgs VEV of order $10 \TEV$ may not be excluded~\cite{Hall:2014dfa}. 
However, 
it has been argued that 
the Higgs VEV cannot be larger than about $1 \TEV$ 
when the Yukawa couplings are fixed~\cite{Agrawal:1997gf}.

\subsection{LSP abundance}

We consider wino or higgsino LSP with a mass lighter than $\mathcal{O}(1) \TEV$ 
because 
otherwise its thermal relic abundance is larger than the observed DM abundance~\cite{Ibe:2012hu, Hall:2012zp, Wells:2004di}. 
Such a light wino or higgsino 
in addition to $100 \TEV$ gravitino 
are naturally realized in pure gravity mediation~\cite{puregm, Ibe:2011aa}. 
In this model, 
the neutral wino acquires one-loop suppressed soft masses through the anomaly 
mediated SUSY breaking effect~\cite{Randall:1998uk, Giudice:1998xp}: 
\beq 
 m_{\tilde{w}} &\simeq& \frac{g_2^2}{16 \pi^2} \lmk \mg + L \rmk 
 \\
 &\simeq& 3 \times 10^{-3} \lmk m_{3/2} + L \rmk 
 \\
 L &\equiv& \mu_H \sin 2 \beta \frac{m_A^2}{\abs{\mu_H}^2 - m_A^2 } \log \frac{\abs{\mu_H}^2}{m_A^2},  
\eeq
where $L$ is Higgsino threshold corrections~\cite{Giudice:1998xp, Gherghetta:1999sw}, 
$m_A$ is the mass of the heavy Higgs bosons, 
$\mu_H$ is the Higgs $\mu$-term, 
and $\tan \beta$ is the ratio of the VEV of $H_u$ and $H_d$. 
Neutral higgsino can also be the LSP when $\mu_H$ is smaller than $m_{\tilde{w}}$. 
The thermal relic abundance of the LSP is subdominant compared with the observed DM abundance 
when 
the LSP is wino with mass lighter than $2.9 \TEV$ 
or higgsino with mass lighter than $1 \TEV$.

Since the gravitino mass is of order $100 \TEV$, 
it decays into radiation before the BBN epoch. 
The produced LSP from gravitino decay 
does not overclose the Universe 
for reheating temperature of order $10^9 \GEV$~\cite{Ibe:2012hu}.

Indirect detection experiments of DM puts lower bounds 
on the LSP abundance. 
If the LSP abundance is equal to that of DM, 
the wino LSP with $m_{\tilde{w}} \le 390 \GEV$ and 
$2.14 \TEV \le m_{\tilde{w}} \le 2.53 \TEV$ 
is excluded~\cite{Bhattacherjee:2014dya} 
and 
the higgsino LSP with $m_{\tilde{h}} \le 160 \GEV$ is 
excluded~\cite{Fan:2013faa}. 
The future indirect detection experiments 
can detect the wino LSP with $m_{\tilde{w}} \le 1.0 \TEV$ and 
$1.66 \TEV \le m_{\tilde{w}} \le 2.77 \TEV$~\cite{Bhattacherjee:2014dya} 
if it is the dominant component of DM. 
However, 
let us emphasize that 
the LSP may be subdominant component of DM 
in the anthropic landscape. 
This is 
because, 
as we explain in the next section, 
the saxion initial amplitude is fine-tuned 
in the anthropic landscape 
such that 
the LSP produced from its decay does not overclose the Universe. 
Note also that the axion 
is another candidate of DM 
and its 
misalignment angle is fine-tuned such that 
its energy density is below that of DM.

\subsection{Discussion on a small parameter in \Kahler potential}
\label{eternal inf}

In this subsection, we 
provide a scenario that explains the smallness of the parameter $k$ in the \Kahler potential. 
Suppose that 
inflation occurs at the GUT scale 
before the new inflation occurs~\cite{Izawa:1996dv}. 
Let us emphasize that the observable universe 
is determined by the new inflation, 
which lasts more than $50$ $e$-foldings, 
so that there is neither the monopole overproduction problem nor the cosmic string constraint. 
For example, 
the GUT-scale inflation can be realized in the following model:%
\footnote{
We assume no mixing between $\phi$ and $S$ in \Kahler potential. 
}
\beq
 W_{\rm GUT} = \kappa S \lmk \psi \bar{\psi} - \mu^2 \rmk. 
\eeq
where $\kappa$ is a parameter and $\mu = E_{\rm GUT} \approx 10^{-2}$. 
When $S$ has a VEV larger than the parameter $\mu$, 
the hybrid inflation occurs~\cite{Copeland:1994vg, Dvali:1994ms}.

The Hubble parameter during the GUT-scale inflation is given by 
\beq 
 H_{\rm GUT} \simeq \kappa \mu^2 / \sqrt{3}. 
\eeq
Through the supergravity effect, 
the inflaton in the new inflation sector obtains 
terms of 
\beq
 V_H (\phi) = c_H H_{\rm GUT}^2 \abs{\phi}^2 - a_H \la W_{\rm GUT} \ra v^2 \phi + c.c. + \dots, 
\eeq
where $\dots$ represents irrelevant higher-dimensional terms~\cite{Dine:1995uk}. 
The Hubble-induced mass term makes $\phi$ stay at the minimum of 
\beq
 \la \phi \ra_{\rm ini} 
 &\simeq& \frac{a_H}{c_H} W_{\rm GUT} \frac{v^2}{H_{\rm GUT}^2} 
 \\
 &\simeq& 3 \frac{a_H v^2}{c_H \kappa \mu}, 
 \label{vev}
\eeq
at the end of the GUT-scale inflation. 
This is also true after the GUT-scale inflation.%
\footnote{
Note that 
$W_{\rm GUT} \simeq \sqrt{2} \kappa \mu S \delta_\psi$ 
around the minimum of the potential, 
where the waterfall field $\delta_\psi$ is defined by $\psi = \bar{\psi} = \mu + \delta_\psi / \sqrt{2}$. 
Since the waterfall field as well as the inflaton $S$ 
start to oscillate after inflation, 
$W_{\rm GUT}$ is in general nonzero and is proportional to $a^{-3} \propto H^2 (t)$. 
Thus, even after the GUT-scale inflation, 
$\la \phi \ra_{\rm ini}$ is given by \eq{vev} 
with $\mathcal{O}(1)$ suppression 
due to a phase difference between $S$ and $\delta_\psi$ oscillations. 
}
Some time after the GUT-scale inflation ends, 
the energy density of the Universe becomes dominated by the F-term of $\phi$ 
and new inflation occurs. 
Equation~(\ref{vev}) gives the initial amplitude of $\phi$ at the beginning of new inflation.

Here let us consider the condition that the new inflation can be eternal inflation~\cite{Guth:1982pn, 
Vilenkin:1983xq, Linde:1986fc}. 
The inflation is eternal when 
the amplitude of 
quantum fluctuations of the horizon scale 
is larger than the classical motion of $\phi$. 
This condition 
is quantitatively written as 
\beq
 \frac{H_{\rm new}^2}{\dot{\phi}} \gtrsim 5, 
\eeq
at the beginning of new inflation. 
Using the slow roll relation of $\dot{\phi} \simeq V_{\rm new}' / 3 H_{\rm new}$, 
we can rewrite the condition as 
\beq
 5 &\lesssim& \frac{3 H_{\rm new}^3}{V_{\rm new}'} 
 \\
 \leftrightarrow 
  \la \phi \ra_{\rm ini} &\lesssim& \frac{v^2}{5\sqrt{3} k}. 
\eeq
Together with Eq.~(\ref{vev}), 
this condition can be rewritten as 
\beq
 k &\lesssim& \kappa \frac{c_H}{15 \sqrt{3} a_H} \mu 
 = \mathcal{O} (1) \times \mu 
 \\
 &\approx& 0.01, 
\eeq
where we use $\mu \approx 10^{-2}$ in the last line. 
Thus 
the eternal inflation occurs 
for $k$ of order $0.01$. 
When the eternal inflation occurs, 
the number of universes 
becomes infinitely large. 
On the other hand, the parameter $k$ should be as large as possible. 
Thus we expect that 
the parameter $k$ is determined 
as the largest value with which the new inflation is eternal. 
This is the reason why 
the parameter $k$ has a value of order $0.01$. 
As a result, 
the spectral index is naturally consistent with the observation [see \eq{n_s}].

\section{SUSY axion cosmology 
\label{sec3}}

The string theory predicts many axions 
with their decay constants of order the GUT scale~\cite{
Conlon:2006tq, Svrcek:2006yi, Choi:2006za}. 
We expect that 
at least one of these axions couples to gluons 
so that the strong CP problem is solved by the Peccei-Quinn (PQ) mechanism~\cite{Peccei:1977hh}. 
This is very interesting 
because the smallness of the strong CP phase cannot be explained in the anthropic landscape. 
In fact, 
we can live in a universe with the strong CP phase $\theta \sim 10^{-2}$, 
which is much larger value than the experimental constraint $\theta \lesssim 10^{-10}$~\cite{Baker:2006ts}. 
Although an axion oscillation with a string-scale decay constant 
may overclose the Universe after the QCD phase transition~\cite{
Preskill:1982cy, Abbott:1982af, Dine:1982ah}, 
this problem can be addressed by the fine-tuning of its initial angle in the anthropic landscape. 
This implies that the axion contributes to DM 
with a fraction of order unity.

In SUSY, 
there is a super-partner of axion, called saxion, 
so that we have to take into account its dynamics in the early Universe. 
The saxion has a mass of order the gravitino mass 
and it starts to oscillate around its low-energy minimum 
when the Hubble parameter decreases to its mass scale. 
Since the initial amplitude of saxion oscillation 
is naively of order the GUT scale,
its energy density may soon overclose the Universe. 
This problem can be addressed by the fine-tuning of its initial amplitude in the anthropic landscape. 
This scenario is similar to the one in our previous work~\cite{Kawasaki:2015pva}, 
where we assumed that 
the R-parity and matter parity are violated 
so that 
the LSP produced from saxion decay 
also decays before the BBN epoch 
in order not to overclose the Universe. 
However, 
the matter parity may be originated from the spontaneous breaking of $U(1)_{B-L}$ symmetry, 
which naturally results in the realization of seesaw mechanism 
to account for the smallness of left-handed neutrino masses~\cite{seesaw}.%
\footnote{
Although the R-parity can be originated from $Z_{10R}$ defined in Table.~\ref{table1}, 
it is not the case in general. 
}
Motivated by this issue, 
in this paper 
we consider the case that 
the R-parity and matter parity is conserved and the LSP is stable. 
In this case, 
the saxion initial VEV is fine-tuned in the anthropic landscape 
so that the LSP produced from saxion decay does not overclose the Universe.%
\footnote{
One might wonder that 
the mass of LSP can be fine-tuned instead of the fine-tuning of saxion initial VEV 
so that its abundance does not overclose the Universe. 
However, 
the fine-tuning of the LSP mass is severer than that of the saxion initial VEV. 
Thus the latter scenario is favoured in the anthropic landscape. 
}
This implies that 
the saxion decay does not produce much entropy 
and hence the baryon number produced via the thermal leptogenesis is not diluted.

\subsection{Saxion decay
\label{sec:saxion}}

As we discuss in the previous section, 
the anthropic landscape implies the gravitino mass of order $100 \TEV$. 
Thus the mass of saxion is also of order $100 \TEV$. 
Here we consider saxion decay 
and check that it decays before the BBN epoch 
not to spoil the success of the BBN theory~\cite{Kawasaki:1999na}.

The saxion decays into gauge bosons and gauginos through 
\beq
 \mathcal{L} \supset \int \dd^2 \theta 
 \frac{\sqrt{2} g^2 A}{32 \pi^2 f_a} W^\alpha W_\alpha + h.c., 
\eeq
where $A$ is an axion chiral superfield 
and $f_a$ is the axion decay constant of order the GUT scale. 
The decay rate of saxion into gluons is given by 
\beq
 \Gamma_g = \frac{\alpha_s^2}{32 \pi^3} \frac{m_\sigma^3}{f_a^2}, 
\eeq
where $m_\sigma$ is saxion mass. 
If kinematically allowed, 
the saxion decays also into gluinos with a rate of 
$\Gamma_{\tilde{g}} \simeq (m_{3/2} / m_\sigma )^2 \Gamma_g$. 
Its decay temperature is therefore 
given by 
\beq
 T_{\rm d} 
 \simeq 11 \MEV \lmk \frac{m_\sigma}{100 \TEV} \rmk^{3/2}
 \lmk \frac{f_a}{10^{16} \GEV} \rmk^{-1}. 
\eeq
This should be larger than $1 \MEV$ not to spoil the success of the BBN theory~\cite{Kawasaki:1999na}. 
The saxion mass of order $100 \TEV$ is consistent with this bound.

\subsection{Axion isocurvature
\label{sec:axion}}

Next we consider the axion isocurvature perturbation problem. 
Since the axion is massless, 
it acquires quantum fluctuations during inflation~\cite{Linde:1984ti,Seckel:1985tj,Lyth:1989pb,Turner:1990uz,Linde:1991km}: 
\beq
 \delta a \simeq \frac{H_{\rm new}}{2 \pi}. 
\eeq
After inflation, 
the axion fluctuations 
lead to isocurvature modes in density perturbations whose amplitude is given by 
\beq
 P_{\rm iso} \simeq 
 \lmk \frac{\Omega_a h^2}{ \Omega_{\rm DM} h^2} \rmk^2 
 \lmk \frac{H_{\rm inf}}{ \pi f_a \theta_{\rm ini}} \rmk^2, 
\eeq
where $\theta_{\rm ini}$ is the axion misalignment angle and $\Omega_a$ is the axion abundance. 
This amplitude is bounded above by the observations of CMB fluctuations~\cite{Ade:2015lrj}: 
\beq
 \beta_{\rm iso} \equiv \frac{P_{\rm iso}}{P_{\rm ad}} \lesssim 0.037, 
\eeq
where the amplitude of adiabatic perturbations is measured such as $P_{\rm ad} \simeq 2.2 \times 10^{-9}$~\cite{Ade:2015lrj}. 
In the anthropic landscape, 
the axion misalignment angle is 
fine-tuned such that 
its oscillation energy density does not overclose the Universe. 
This implies that 
$\theta_{\rm ini} \lesssim 0.003$ for $f_a \simeq 10^{16} \GEV$~\cite{Preskill:1982cy, Abbott:1982af, Dine:1982ah}
and 
its energy density contributes an $O(1)$ fraction of total DM density in our Universe. 
Thus the isocurvature constraint implies that 
the energy scale of inflation is bounded above 
such as 
\beq
 H_{\rm new} \lesssim 9.6 \times 10^8 \GEV 
 \lmk \frac{\Omega_a h^2}{0.12} \rmk^{-1/2}
 \lmk \frac{f_a}{10^{16} \GEV} \rmk^{0.4}. 
 \label{H_inf bound}
\eeq
where we use $\Omega_{\rm DM} h^2 \simeq 0.12$. 
This is satisfied in the new inflation model 
considered in the previous section.

Note that the axion isocurvature constraint cannot be avoided directly 
in the anthropic landscape  
because 
our Universe 
can exist with an $O(1)$ fraction of isocurvature modes. 
Low-scale inflation is implied by the string landscape 
and the new inflation model is a natural realization of such low scale inflation.

\section{Conclusion 
\label{sec:conclusion}}

We consider a new inflation model 
where R-symmetry is broken at the minimum of inflaton potential. 
The reheating temperature is related to the VEV of the inflaton, 
so that the gravitino mass, which is proportional to the R-symmetry breaking effect, 
is then related to the reheating temperature. 
As a result, the realization of thermal leptogenesis 
requires gravitino mass heavier than $100 \TEV$. 
This is consistent with the present constraint on SUSY breaking scale 
and the $125 \GEV$ Higgs boson. 
Furthermore, the new inflation consistently predicts 
the amplitude of density perturbations. 
The spectral index can be consistent with the observed value 
when we allow a fine-tuning in the \Kahler potential. 
This fine-tuning could be explained 
once we require eternal inflation in the new inflation.

Other cosmological problems are also addressed in this scenario 
by the heavy gravitino and the anthropic landscape. 
String axions do not overclose the Universe by the fine-tuning of their initial amplitudes 
in the anthropic landscape. 
The axion isocurvature fluctuations are suppressed 
because the energy scale is sufficiently low in the new inflation model. 
Saxions are so heavy that they decay into LSP before the BBN epoch 
because the gravitino mass is of order $100 \TEV$. 
We assume R-parity conservation 
so that 
the saxion initial amplitude is fine-tuned such that 
its decay products do not overclose the Universe. 
This implies that 
saxions never dominate the Universe 
and hence the saxion decays do not produce a large entropy, 
which is crucial for the successful thermal leptogenesis. 
The string landscape implies that 
DM is the mixture of axion and LSP (wino or higgsino). 
This means that 
the constraint on wino DM by the indirect DM search experiments 
can be avoided even if the mass of wino is so small that 
we can find it by the LHC experiment with $13 \TEV$ center of energy.

\vspace{1cm}

%
\section*{Acknowledgments}
T.\,T.\,Y. thanks Prof. Raymond Volkas for the hospitality
during his stay at the University of Melbourne.
This work is supported by Grants-in-Aid for Scientific Research from the Ministry of Education, Culture, Sports, Science, and Technology (MEXT), Japan,
No. 25400248 (M.\,K.)
and  No. 26104009 (T.\,T.\,Y.); 
MEXT Grant-in-Aid for Scientific Research on Innovative Areas No.15H05889 (M.\,K.);
Grant-in-Aid No. 26287039 (T.\,T.\,Y.) from the Japan Society for the Promotion of Science (JSPS); 
World Premier International Research Center Initiative (WPI), MEXT, Japan (M.\,K., M.\,Y., and T.\,T.\,Y.), 
and the Program for the Leading Graduate Schools, MEXT, Japan (M.Y.).
M.Y. acknowledges the support by the JSPS Research Fellowships for Young Scientists, No. 25.8715.
The research leading to these results has received funding
from the European Research Council under the European Unions Seventh
Framework Programme (FP/2007-2013) / ERC Grant Agreement n. 279972
``NPFlavour'' (N.\,Y.).
%

\vspace{1cm}



\end{document}